\documentclass[11pt,a4paper,reqno]{amsart}
\title{Lie Bi-Algebras on the Non-Commutative Torus}
\author{Giovanni Landi and S. G. Rajeev}
\date{June 2021}
\usepackage{amsmath,amsthm,amsfonts}
\usepackage{amssymb}
\usepackage{amstext}
\usepackage{mathtools}
\usepackage{graphicx}
\usepackage[english]{babel}
\usepackage{fullpage}
\usepackage{hyperref}

\numberwithin{equation}{section}

\newtheorem{theorem}{Theorem}[section]
\newtheorem{lemma}[theorem]{Lemma}

\theoremstyle{definition}
  
\newtheorem{example}[theorem]{Example}

\newcommand{\A}{\mathcal{A}}

\newcommand{\C}{\mathbb{C}}
\newcommand{\R}{\mathbb{R}}

\newcommand{\T}{\mathbb{T}}
\newcommand{\Z}{\mathbb{Z}}
\newcommand{\IS}{\mathbb{S}}

\newcommand{\fb}{\mathfrak b}

\newcommand{\fh}{\mathfrak h}

\newcommand{\fn}{\mathfrak n}
\newcommand{\ft}{\mathfrak t}

\newcommand{\hs}[2]{\langle #1, #2 \rangle}

\def\bra#1{\left\langle #1\right|}
\def\ket#1{\left| #1\right\rangle}
\def\hs#1#2{\left\langle #1,#2\right\rangle}

\newcommand{\tr}{\textnormal{Tr}}
\newcommand{\ii}{\mathrm{i}}

\newcommand{\wt}{\widetilde}

\newcommand{\beq}{\begin{equation}}
\newcommand{\eeq}{\end{equation}}

\newcommand{\nn}{\nonumber}

\newcommand{\ignore}[1]{}
\def\u#1{\underline{#1}}

\begin{document}

\address[Giovanni Landi]{Matematica, Universit\`a di Trieste, Via A. Valerio 12/1, 34127 Trieste, Italy,
Institute for Geometry and Physics (IGAP) Trieste, Italy, 
and INFN, Sezione di Trieste, Trieste, Italy. 
landi@units.it}
\address[S. G. Rajeev]{
Department of Physics and Astronomy,
Department of Mathematics,
University of Rochester,
Rochester,NY 14627,
USA.
s.g.rajeev@rochester.edu}

\begin{abstract}
Infinitesimal symmetries of a classical mechanical system are usually described by a Lie algebra acting on the phase space, preserving the Poisson brackets. We propose that a quantum analogue is  the action  of a Lie bi-algebra on the associative $*$-algebra of observables. The latter can be thought of as functions on some underlying non-commutative manifold. We illustrate this for the   non-commutative torus $\mathbb{T}^2_\theta$. The canonical trace defines a Manin triple from which a Lie bi-algebra can be constructed. In the special case of rational $\theta=\frac{M}{N}$ this Lie bi-algebra is $\u{GL}(N)=\u{U}(N)\oplus \u{B}(N)$, corresponding to unitary and upper triangular matrices. The Lie bi-algebra has a remnant  in the classical limit $N\to\infty$:  the elements of $\u{U}(N)$ tend to real functions while $\u{B}(N)$ tends to a space of  complex analytic functions.
\end{abstract}

\maketitle

\tableofcontents

\parskip = .75 ex

\section{Introduction}
In the quantum theory the phase space cannot be a manifold in the usual sense, since momentum and position cannot be simultaneously measured. The appropriate generalization  needed to describe the quantum phase space is  non-commutative geometry. The algebra of quantum observables is an associative algebra which tends in the classical limit to the  Poisson algebra of functions on the classical phase space; that is, a commutative algebra to order zero in $\hbar$ along with a Poisson bracket that gives the correction  of order $\hbar$.

Topological properties of the manifold can be encoded into properties of the algebra of functions on it, which then carry over to the non-commutative case, as properties of the quantum observable algebra. There is a well-developed K-theory of operator algebras \cite{KThOpAlg}:  familiar invariants such as Chern numbers  have non-commutative counterparts. 

A third theme is that of symmetry. The continuous symmetries of a classical system form a Lie group  which acts on the classical phase space. In simple cases, this Lie group  extends to the quantum theory, with a representation on the Hilbert space of states. Sometimes, quantization modifies the symmetry. For example, the  symmetry of the quantum theory can be the central extension of the classical one (the corresponding representation is a projective representation of the classical symmetry). Such deformations are called ``anomalies" in the physics literature. In the most well-known case, the conformal anomaly leads to the Virasoro algebra  \cite{Mickelsson}. In other examples the quantum symmetry is a new Lie group, of which the classical limit is a Wigner contraction. 

An even more general possibility is that the symmetries of the quantum theory do not form a  group at all.  Instead, it  is a ``quantum group". More precisely, it  is a Hopf algebra with an action (or co-action) on the algebra of quantum observables. Again, there is a classical remnant of this phenomenon: there is a Poisson bracket on the group itself, such that the group multiplication is a Poisson map \cite{ChariPressley}. Such a Poisson-Lie group $A$ acts  on the classical  phase space $M$, such that  the induced co-action $C(M)\to C(M)\otimes C(A)$ is a Poisson map. If the Poisson bracket on $A$ is zero, this reduces to the familiar condition that the group action leave the Poisson bracket on $M$ invariant; the generating function $f:M\to \u{A}'$  of the infinitesimal action  is the ``moment map", valued in the dual $\u{A}'$  of the  Lie algebra $\u{A}$ of $A$. More generally, the moment map is a function $f:M\to A'$ valued in a Lie group dual of $A$.

 Infinitesimally, a Poisson-Lie group is a Lie bi-algebra. The infinitesimal group multiplications give the Lie bracket as usual. In addition, the Poisson bracket on the group $A$ becomes a co-product on the Lie algebra $\u{A}$. The  product and co-product must satisfy an infinitesimal version of the condition that the group multiplication be a Poisson map.

These more general ideas about symmetries have not yet been fully utilized in physics, except in the narrow   context of integrable systems. They  have  emerged recently in the very chaotic case of incompressible fluid mechanics \cite{RajeevHelicity} (Euler equation of an ideal fluid).  It would be interesting to have more examples of physical systems with quantum group or Lie bi-algebra symmetry.

 We study  the simplest case of a phase space: a two dimensional torus. The quantum geometry is that of the non-commutative two-torus  
 $\T^2_\theta$.
 We uncover a Lie bi-algebra hidden in $\T^2_\theta$. More precisely, for the algebra of function on $\T^2_\theta$, that is the complex associative $*$-algebra generated by two unitary elements $P,Q$ satisfying the relations
\beq
 	PQ=\omega \, QP, \qquad \omega=e^{2\pi \ii \theta}	.
\eeq
The commutator of two elements $[F,G]=FG-GF$ yields a Lie algebra $\u{S}_\theta $. The canonical invariant trace $\tau$ 
on $\T^2_\theta$ yields an invariant inner product
\beq
\langle F,G\rangle = \tau (FG)
\eeq
for this Lie algebra.  This inner product is not positive. In fact there are subspaces $\u{A}$ and $\u{B}$ on which it vanishes such that 
$\u{S}_\theta=\u{A}\oplus \u{B}$; then $\u{A}$ and $\u{B}$ are dual to each other as vector spaces. That is,  $(\u{S}_\theta,\u{A},\u{B})$ is a Manin triple. Another point of view is that $\u{A}$ has a co-product (the mirror image of the Lie bracket in $\u{B}$) which turns it into a Lie bi-algebra.

 As $\theta\to 0$ this complex associative $*$-algebra tends to the Poisson algebra of complex-valued functions on 
 the torus $\T^2$ with the bracket
\beq
\{F,G\}={\partial F\over \partial x_1}{\partial G \over \partial x_2}-{\partial F\over \partial x_2}{\partial F\over \partial x_1}
\eeq
and identification  ${P=e^{2\pi  \ii x_1},Q=e^{2\pi  \ii x_2}}$. We will see that this also is an (infinite dimensional) Lie bi-algebra. The invariant trace tends to the integral, yielding the invariant inner product
\beq
\langle F,G\rangle =\mathrm{Im} \int FG\  {dx^1dx^2\over (2\pi)^2} .
\eeq

The classical limit  of $\u{A}$ is the sub-space of real-valued functions; the elements  of the classical limit of $\u{B}$ are complex-valued functions on the torus allowing an analytic continuation to the interior of the unit disc in the variable $P$ (and satisfy a reality condition when independent of $P$). This Lie bi-algebra (for the Poisson structure) is the classical remnant of the quantum group symmetry of the non-commutative torus.

The physical meaning of $\u{A}\subset \u{S}_\theta$  is clear: it is the quantum counterpart to the Lie algebra of canonical transformations. The corresponding group $A$  is the familiar group of unitary transformations. The physical meaning of $\u{B}$  (or equivalently, the co-product on $\u{A}$)  remains somewhat mysterious. The co-product is the structure that allows us to combine two representations of $\u{A}$ into a new one. In the simplest case (when the co-product is trivial) this is just the direct sum of representations. The surprise is that there are more general ways  of combining representations, given the co-product. Physically, this means that there are new ways of combining two independent quantum systems to get a new ones, while preserving the invariance under unitary transformations. This ought to be important in the context of quantum computation, where the plan is to combine may small quantum systems to get a large processor.

What is the quantum group (Hopf algebra)  associated to the Lie bi-algebra $\u{S}_\theta$ ? Finite dimensional Lie bi-algebras can be exponentiated into quantum groups, as conjectured by Drinfeld (and Jimbo) originally. There is no such general construction in the infinite dimensional cases. We must approach this with a ``regularization": approximate the infinite dimensional algebra by a sequence of finite dimensional ones, and then look at the limit as the dimension tends to infinity. A prototype of this is the approximation of real (irrational) numbers by rational ones.

So we look at the 
 the case of  rational  $\theta={M\over N}$ (i.e.,  $\omega$ is a primitive $N$th root of unity). Then we can impose the additional relations $P^N=1=Q^N$, giving a finite dimensional algebra.
There is then a simple explicit realization using clock-shift operators ($j$ is taken modulo $N$)
\beq\label{eq:RatNCTorus}
P=\sum_{j=1}^N \ket{j+1}\bra{j}, \qquad Q=\sum_{j=1}^N \omega^j \ket{j}\bra{j} .
\eeq
If $N=2$ these are Pauli matrices (acting on ``qubit" states), while for $N=3$ they can be realized in terms of Gell-Mann matrices (acting on ``qutrit" states). More generally, the commutators of  ${P^a Q^b}$ close on  the $N^2$-dimensional complex algebra $gl_N$; in this basis, it is the ``sine-algebra". There is a natural Lie bi-algebra structure on this $gl_N$. That is,  a splitting  $gl_N=\u{A}\oplus \u{B}$ into sub-algebras $\u{A}$ and $\u{B}$ which are isotropic and dual to each other w.r.t. an invariant inner product (a ``Manin triple"). A natural choice is $\u{A}=u_N$ (anti-hermitian matrices) and $\u{B}=sb_N$,  consisting of upper triangular matrices with real entries along the diagonal. It is known that this Lie algebra is the infinitesimal version of the well-known \cite{ChariPressley}  quantum group $SL_\omega(N)$. There is already a well developed representation theory of this Hopf algebra, which should have interesting physical consequences.

 The operators of the rational non-commutative torus defined in \eqref{eq:RatNCTorus} appear, surprisingly, in experimental realizations of ``Qudit" processors \cite{Bloketal}. It is possible that the non-commutative torus provides an alternative model of quantum computation, instead of arrays of ``qubits".  Maintaining quantum coherence for many qubits is experimentally challenging. It might be easier to use instead a smaller number of ``Qudit" systems, each one a rational NC torus for  some $N>2$. Other intriguing connections between operators algebras and complexity theory of computing are also emerging recently, cf. \cite{VidickNoticesAMS}.

Perhaps this theme plays out more generally:  the analogue of diffeomorphisms for non-commutative geometries  could be quantum groups; that is, there is a Hopf algebra co-action on the associative algebra of the non-commutative manifold. As a prototype of this, we recall in Appendix \ref{sec:Taft} the co-action of the Taft-Hopf algebra on the non-commutative torus.  In the classical limit we don't just get commutative algebras and Lie groups; there is in addition a Poisson bracket with a Lie bi-algebra of symmetries.

The non-commutative torus also gives us an example of how topological properties (via $K$-theory) of the classical phase space lift to the quantum theory. It is well-known that  $K_0(\T^2)=\Z^2$;  vector bundles on a torus are classified by a pair of integers which are the  rank and Chern number. This continues to be true in  the non-commutative case: operator theoretic $K$-group $K_0(\T^2_\theta)$ is also $\Z^2$. 
For rational $\theta$ this can be understood in terms of Morita equivalence of vector bundles over $\T^2_\theta$ to those over $\T^2$, while for irrational $\theta$ the classes can be realised in terms of Heisenberg modules \cite{KTheoryNonCommGeometry}. 
But for irrational $\theta$ we have some additional structure. The trace on $\T^2_\theta$ 
embeds $\Z^2\to \mathbb{R}$ as a subgroup of the real numbers:
\[
\tau:(r,m)\mapsto r+m\theta .
\]
In particular, it turns $K_0(\T^2_\theta)$ into an ordered group \cite{RudinFourierGroups}. 
In the limit $\theta\to 0$ the embedding above is lost. 
Remarkably, $K_0(\T^2)$ is still an ordered group with a non-archimedean order. We will use such a lexicographic 
order in constructing a Lie bi-algebra structure for the Poisson algebra on the classical commutative torus while 
for a corresponding Lie bi-algebra structure for the non-commutative torus we use the order of $K_0(\T^2_\theta)$, 
which is most natural for $\T^2_\theta$. In the limit the latter ordering tends to the lexicographic one, as seen in Fig.\ref{fig:orderedgroupZ2}.
The classical remnant of non-commutative geometry contained in the Poisson algebras needs to be studied further.

\section{ Manin triples and Lie bi-algebras}
We recall the main definitions \cite{ChariPressley}.  

 A \emph{Lie bi-algebra}  is a  product  $\Gamma:\u{A}\otimes \u{A}\to \u{A} $ along with a co-product $\Delta:\u{A}\to \u{A}\otimes \u{A}$ 
\[
\Gamma(X_a,X_b)\equiv \left[X_{a},X_{b}\right]=\Gamma_{ab}^{d}X_{d}, \qquad\Delta(X_{a})=\Delta_{a}^{cd}X_{c}\otimes X_{d}
\]
satisfying antisymmetry, $\Gamma_{ab}^d=-\Gamma_{ba}^d$ and $\Delta^{ab}_d=-\Delta^{ba}_d$,  
the Jacobi and co-Jacobi identities
\beq\label{eq:Jacobi}
\Gamma_{ab}^d\Gamma_{dc}^e+\Gamma_{bc}^d\Gamma_{da}^e+\Gamma_{ca}^d\Gamma_{db}^e=0
\eeq
\beq\label{eq:co-Jacobi}
\Delta^{ab}_d\Delta^{dc}_e+\Delta^{bc}_d\Delta^{da}_e+\Delta^{ca}_d\Delta^{db}_e=0
\eeq
and the compatibility condition that $\Delta$ be an infinitesimal automorphism of $\Gamma$:
\[
\left[\Delta\left(X_{a}\right),X_{b}\right]+\left[X_{a},\Delta\left(X_{b}\right)\right]=\Gamma_{ab}^{c}\Delta\left(X_{c}\right).
\]
The latter  amounts to the condition that  $\Delta$ is  closed in the Lie algebra cohomology $H^1(\u{A},\u{A}\otimes \u{A})$:
\beq\label{eq:Cohom}
\left(\partial \Delta \right)^{be}_{ac}\equiv \left[\Gamma_{ad}^{b}\Delta_{c}^{de}+\Gamma_{ad}^{e}\Delta_{c}^{bd}-a\leftrightarrow c\right]-\Gamma_{ac}^{d}\Delta_{d}^{be}=0
\eeq

 A more familiar way to state these conditions is that $\u{S}=\u{A}\oplus \u{B}$ (where $\u{B}$ is the dual vector space of $\u{A}$) is itself a Lie algebra;  in a basis  $X_a\in \u{A}$  and the dual basis $X^a \in \u{B} $ we must have 
 \beq\label{eq:LieBiAlg1}
[X_{a},X_{b}]=\Gamma_{ab}^{c}X_{c}, \qquad
[X^{a},X^{b}]=\Delta_{c}^{ab}X_{c},
\eeq
\beq\label{eq:LieBiAlg2}
 [X^{a},X_{b}]=\Gamma_{bd}^{a}X^{d}-\Delta_{b}^{ad}X_{d}
\eeq
The Jacobi identities from the first line \eqref{eq:LieBiAlg1} are the conditions \eqref{eq:Jacobi}-\eqref{eq:co-Jacobi} above; the mixed Jacobi identities give \eqref{eq:Cohom}. The duality of $\u{A}$ and $\u{B}$ yield an invariant inner product on $\u{S}$:
\[
 \langle X_a,X_b\rangle=0=\langle X^a,X^b\rangle,\quad \langle X_a,X^b\rangle=\delta_a^b.
\]

  A \emph{ Manin triple} is a Lie algebra $\u{S}$ along with Lie sub-algebras $\u{A},\u{B}$ such that $\u{S}=\u{A}\oplus \u{B}$ as vector spaces; moreover $\u{S}$ has an invariant inner product which vanishes when restricted to $\u{A}$ or $\u{B}$ (i.e., $\u{A}$ and $\u{B}$ are isotropic sub-spaces). 
If $\u{S}$ is finite dimensional, every Manin triple admits \cite{ChariPressley} a basis satisfying \eqref{eq:LieBiAlg1}-\eqref{eq:LieBiAlg2}; the notions of Manin triple and Lie bi-algebras are equivalent. However, in the infinite dimensional case, we cannot  rely on  this theorem; we must explicitly verify the commutation relations to obtain a Lie bi-algebra from a Manin triple.

\begin{example}
An example is the Lorentz Lie algebra $\u{SL}(2,C)$ of traceless $2\times 2$ matrices with complex entries (but viewed as a {\em real}  Lie algebra). There is an invariant inner product
\beq
\langle U,V\rangle =\mathrm{Im\ Tr\ }UV .
\eeq
The Lie subalgebra $\u{SU}(2)$ of anti-hermitian traceless matrices is isotropic and can be chosen as $\u{A}$. The complementary space $\u{B}$ cannot be the space of hermitian matrices, as it is not a Lie sub-algebra. Instead we can chose $\u{B}$ to be $\u{SB}(2,C)$, the Lie sub-algebra of traceless upper triangular matrices with real entries along the diagonal. That is,
\[
\u{SL}(2,C)=\u{A}\oplus \u{B}
\]
where
\[
 \u{A}=\u{SU}(2)=\left\{\begin{pmatrix} \ii a&b\\ -\bar{b} & -\ii a \end{pmatrix}\mid a\in \mathbb{R}, b\in\mathbb{C}\right\}
 \]
 \[
\u{B}=\u{SB}(2,C)=\left\{\begin{pmatrix}a&b\\ 0 & -a \end{pmatrix}\mid a\in \mathbb{R}, b\in\mathbb{C}\right\}\
 \]
is a Manin triple.
A basis for $\u{A}$ can be built out of Pauli matrices, $X_a=-{i\over 2}\sigma_a$:
\[
X_1=\left(
\begin{array}{cc}
 0 & -\frac{\ii}{2} \\
 -\frac{\ii}{2} & 0 \\
\end{array}
\right),\quad
X_2=\left(
\begin{array}{cc}
 0 & -\frac{1}{2} \\
 + \frac{1}{2} & 0 \\
\end{array}
\right),\quad
X_3=\left(
\begin{array}{cc}
 - \frac{\ii}{2} & 0 \\
 0 & \frac{\ii}{2} \\
\end{array}
\right)
\]
which determines the  dual basis for $\u{B}$:
\[
X^1=\left(
\begin{array}{cc}
 0 & -2 \\
 0 & 0 \\
\end{array}
\right),\quad
X^2=\left(
\begin{array}{cc}
 0 & 2  \ii \\
 0 & 0 \\
\end{array}
\right),\quad
X^3=\left(
\begin{array}{cc}
 -1 & 0 \\
 0 & 1 \\
\end{array}
\right)
\]
They satisfy 
\[
\langle X_a,X_b\rangle=0=\langle X^a,X^b\rangle,\langle X_a,X^b\rangle=\delta_a^b
\]
The structure constants $\Gamma$ are the familiar ones from angular momentum theory
\[
\Gamma^1_{23}=1=\Gamma^2_{31}=\Gamma^3_{12}
\]
along with 
\[
 \Delta^{23}_2=2=\Delta^{13}_1.
\]
The remaining components of $\Gamma,\Delta$ are zero, unless related to these by anti-symmetry. 
Also, they satisfy the compatible commutation relations \eqref{eq:LieBiAlg2} above.
\end{example}

\begin{example}
Next, consider the Lie algebra $\u{SL}(3,\C)$ of traceless $3\times 3$ matrices with complex entries 
(again, viewed as a real  Lie algebra). 
The subalgebra $\u{SU}(3)$ of anti-hermitian\footnote{
 We use a compromise between the physics and mathematical conventions for Lie algebras.  For matrix Lie algebras $\u{SU}(N)$, we use  the mathematical convention that they consist of anti-hermitian matrices, whose commutators are also anti-hermitian. This differs by a factor $i$ from the physics convention \cite{Georgi}. In the classical limit (Section \ref{sec:ClassicalLimit}),  we use the physics convention that $\u{SU}(N)$ elements tend to real functions on the torus.}
traceless matrices is again isotropic and is dual to 
$\u{SB}(3,\C)$, the sub-algebra of upper triangular traceless matrices with real entries along the diagonal, which is  isotropic as well.
\[
\u{SL}(3,\C)=\u{SU}(3) \oplus \u{SB}(3,\C)
\]
 We take a basis of
for $\u{SU}(3)$ in terms of anti-hermitian Gell-Mann matrices \cite{Georgi}, $X_a = -  \ii \lambda_a$:
\[
X_{1}= 
\begin{pmatrix}
0 & -\ii & 0\\
-\ii  & 0 & 0 \\
0 & 0 & 0
\end{pmatrix}, \quad 
X_{2}= 
\begin{pmatrix}
0 & -1 & 0\\
1  & 0 & 0 \\
0 & 0 & 0
\end{pmatrix}, \quad
X_{3}= 
\begin{pmatrix}
-\ii & 0 & 0\\
0 &  \ii & 0 \\
0 & 0 & 0
\end{pmatrix},
\]
\[
X_{4}= 
\begin{pmatrix}
0 & 0 & -\ii \\
0  & 0 & 0 \\
-\ii & 0 & 0
\end{pmatrix}, \quad 
X_{5}= 
\begin{pmatrix}
0 & 0& -1\\
0 & 0 & 0 \\
1 & 0 & 0
\end{pmatrix}, \quad
X_{6}= 
\begin{pmatrix}
0 & 0 & 0\\
0 & 0 & -\ii \\
0 & -\ii  & 0
\end{pmatrix},
\]
\[
X_{7}= 
\begin{pmatrix}
0 & 0 & 0\\
0 & 0 & -1 \\
0 & 1  & 0
\end{pmatrix}, \quad
X_{8}= 
\frac{1}{\sqrt{3}}
\begin{pmatrix}
-\ii & 0 & 0\\
0 & -\ii & 0 \\
0 & 0 & 2 \ii
\end{pmatrix},
\]
A dual basis for $\u{SB}(3,\C)$ is worked out to be
:\[
X^{1}= 
\begin{pmatrix}
0 & -1 & 0\\
0  & 0 & 0 \\
0 & 0 & 0
\end{pmatrix}, \quad 
X^{2}= 
\begin{pmatrix}
0 &  \ii & 0\\
0  & 0 & 0 \\
0 & 0 & 0
\end{pmatrix}, \quad
X^{3}= \frac{1}{2}
\begin{pmatrix}
-1 & 0 & 0\\
0 & 1 & 0 \\
0 & 0 & 0
\end{pmatrix},
\]
\[
X^{4}= 
\begin{pmatrix}
0 & 0 & -1 \\
0  & 0 & 0 \\
0 & 0 & 0
\end{pmatrix}, \quad 
X^{5}= 
\begin{pmatrix}
0 & 0&  \ii \\
0 & 0 & 0 \\
0 & 0 & 0
\end{pmatrix}, \quad
X^{6}= 
\begin{pmatrix}
0 & 0 & 0\\
0 & 0 & -1 \\
0 & 0 & 0
\end{pmatrix},
\]
\[
X^{7}= 
\begin{pmatrix}
0 & 0 & 0\\
0 & 0 &  \ii \\
0 & 0 & 0
\end{pmatrix}, \quad
X^{8}= 
\frac{1}{2 \sqrt{3}}
\begin{pmatrix}
-1 & 0 & 0\\
0 & -1 & 0 \\
0 & 0 & 2 
\end{pmatrix}.
\]
The bases are isotropic and dual to each other: 
\[
\langle X_a, X_b \rangle=0, \quad \langle X^a, X^b \rangle=0, \quad 
\langle X_a, X^b \rangle=\delta_a^b \qquad a, b\in\{1, \dots, 8\} .
\]
The structure constants of the Gell-Mann matrices $[X_a, X_b] = \Gamma_{ab}{}^{c} X_c$ are well known: 
they are completely antisymmetric in the three indices and explicitly given by
\[
\Gamma_{123}=2\ ,\quad \Gamma_{147}=\Gamma_{165}=\Gamma_{246}=\Gamma_{257}=\Gamma_{345}=\Gamma_{376}=1 \ ,\quad \Gamma_{458}=\Gamma_{678}= \sqrt{3} \ .
\]

\noindent
As for the structure constants of the dual basis,  $[X^a, X^b] = \Delta^{ab}{}_{c}X^c$, the non zero ones are:
\begin{align*}
\Delta^{13}{}_{1} & = \Delta^{23}{}_{2} = \Delta^{61}{}_{4} = \Delta^{71}{}_{5} = 
\Delta^{62}{}_{5} = \Delta^{27}{}_{4} = 1, \\
\Delta^{43}{}_{4} & = \Delta^{53}{}_{5} = \Delta^{36}{}_{6} = \Delta^{37}{}_{7} = \frac{1}{2}, \\
\Delta^{84}{}_{4} & = \Delta^{85}{}_{5} = \Delta^{86}{}_{6} = \Delta^{87}{}_{7} = \frac{\sqrt{3}}{2} ,
\end{align*}
and their antisymmetric ones in the two upper indices. They may be  verified to satisfy the compatibility 
condition $[X^{a}, X_{b}]=\Gamma_{bd}{}^{a} X^{d}-\Delta^{ad}{}_{b} X_{d}$. 

Alternative bases can be constructed out of clock and shift matrices in \eqref{eq:RatNCTorus}, which allows for a generalization to arbitrary  dimension $N$ in the following.  Let $\omega = e^{2 \pi  \ii / 3}$, a third root of unity, with 
$1+\omega+\omega^2=0$, $\bar\omega=\omega^2$, $\omega - \omega^2 =  \ii \sqrt{3}$. Then,
a basis for $su(3)$ is given by: 
\[
X_{0}=  \ii (Q + Q^2) =  \ii 
\begin{pmatrix}
2 & 0 & 0\\
0 & -1 & 0 \\
0 & 0 & -1
\end{pmatrix}, \quad 
\wt{X}_{0}= Q - Q^2 =  \ii \lambda
\begin{pmatrix}
0 & 0 & 0\\
0 & 1 & 0 \\
0 & 0 & -1
\end{pmatrix}, 
\]
\[
X_{1}=  \ii (P + P^2) =  \ii 
\begin{pmatrix}
0 & 1 & 1\\
1 & 0 & 1 \\
1 & 1 & 0
\end{pmatrix}, \quad 
\wt{X}_{1}= P - P^2 =  
\begin{pmatrix}
0 & 1 & -1\\
-1 & 0 & 1 \\
1 & -1 & 0
\end{pmatrix},
\]
\[
X_{2}=  \ii (PQ + \omega^2 P^2 Q^2) =  \ii 
\begin{pmatrix}
0 & \omega & 1\\
\omega^2 & 0 & \omega^2 \\
1 & \omega & 0
\end{pmatrix}, \quad 
\wt{X}_{2}= PQ - \omega^2 P^2 Q^2 = 
\begin{pmatrix}
0 & \omega & -1\\
- \omega^2 & 0 & \omega^2 \\
1 & -\omega & 0
\end{pmatrix}, 
\]
\[
X_{3}=  \ii (PQ^2 + \omega P^2 Q) =  \ii 
\begin{pmatrix}
0 & \omega^2 & 1\\
\omega & 0 & \omega \\
1 & \omega^2 & 0
\end{pmatrix}, \quad 
\wt{X}_{3}= PQ^2 - \omega^2 P^2 Q = 
\begin{pmatrix}
0 & \omega^2 & -1\\
- \omega & 0 & \omega\\
1 & -\omega^2 & 0
\end{pmatrix}. 
\]
One verifies directly that this basis is isotropic. These are not Gell-Mann matrices.

A dual basis for $\u{SB}(3,\C)$ can be found using $Q$ and a matrix $R=P- \ket{2} \bra{0}$ satisfying
\[
RQ=\omega \, QR,\quad R^3=0 .
\]
Then, \[
X^{0}= \frac{1}{6}(Q + Q^2) = 
\frac{1}{6}
\begin{pmatrix}
2 & 0 & 0\\
0 & -1 & 0 \\
0 & 0 & -1
\end{pmatrix}, \quad 
\wt{X}^{0}=  \frac{\ii}{2 \lambda^2}(Q^2 - Q) = \frac{1}{2\lambda} 
\begin{pmatrix}
0 & 0 & 0\\
0 & 1 & 0 \\
0 & 0 & -1
\end{pmatrix}, 
\]

\[
X^{1}= \frac{1}{3} (R + R^2) = \frac{1}{3}
\begin{pmatrix}
0 & 1 & 1\\
0 & 0 & 1 \\
0 & 0& 0
\end{pmatrix}, \quad 
\wt{X}^{1}= \frac{\ii}{3} (R - R^2) =\frac{\ii}{3} 
\begin{pmatrix}
0 & -1 & 1\\
0 & 0 & -1 \\
0 & 0 & 0
\end{pmatrix},
\]
\[
X^{2}= \frac{1}{3} (RQ + \omega^2 R^2 Q^2) = \frac{1}{3}
\begin{pmatrix}
0 & \omega & 1\\
0 & 0 & \omega^2 \\
0 & 0 & 0
\end{pmatrix}, \quad 
\wt{X}^{2}= \frac{\ii}{3} (RQ - \omega^2 R^2 Q^2) = \frac{\ii}{3} 
\begin{pmatrix}
0 & -\omega & 1\\
0 & 0 & - \omega^2 \\
0 & 0 & 0
\end{pmatrix}, 
\]
\[
X^{3}= \frac{1}{3} (RQ^2 + \omega R^2 Q) = \frac{1}{3}
\begin{pmatrix}
0 & \omega^2 & 1\\
0 & 0 & \omega \\
0 & 0 & 0
\end{pmatrix}, \quad 
\wt{X}^{3}= \frac{\ii}{3} (RQ^2 - \omega^2 R^2 Q) = \frac{\ii}{3} 
\begin{pmatrix}
0 & -\omega^2 & 1\\
0 & 0 & -\omega\\
0 & 0 & 0
\end{pmatrix}. 
\]
Again one verifies that the basis is isotropic. As for cross pairings one verifies that
\[
\langle X_A, X^B \rangle=\delta_A^B \qquad A, B\in\{0,1,2,3, \wt0,\wt1,\wt2,\wt3\} 
\]
thus the bases are dual to each other.
\end{example}

\section{The rational non-commutative torus}\label{rnct}
%\subsection{The sine-algebra}
We are ready to generalise the construction above out of of clock and shift matrices.
\subsection{Clock and shift matrices}
Let $\omega(=e^{2 \pi \ii / N})$ be a primitive $N$-the root of unity. This implies that 
$\sum_{j=0}^{N-1} \omega^{jm}= N \delta(m)$. 
Consider clock and shift matrices $Q$ and $P$:
\beq\label{csm}
Q^j \ket{m} = \omega^{km} \ket{m} \qquad P^k \ket{m} = \ket{m-k}
\eeq
 and indices $m,k = 0, 1, \cdots, N-1$, defined modulo $N$. 
These matrices obey the relation 
$$
P Q = \omega \, Q P ,
$$
are unitary,  
traceless and  $Q^N = P^N =1$.
We shall also need the truncated matrix
$$
R = P - \ket{N-1}\bra{0}
$$
which is a strictly upper triangular matrix such that:
$$
R Q = \omega \, Q R , \qquad R^N = 0.
$$
The matrices
\beq\label{gencs}
e_{r,s} = \omega^{- \frac{1}{2} rs} P^r Q^s = \omega^{- \frac{1}{2} rs}
P^r \, \sum_{m=0}^{N-1} \omega^{ms} \ket{m}\bra{m} 
%\sum_{m=0}^{N-1} \ket{m-r} \omega^{ms} \bra{m}
\eeq
generate the $N^2$-dimensional complex algebra $\u{GL}_N$. They close on the sine-algebra. Indeed,
$$
e_{j,k} \, e_{r,s} = \omega^{\frac{1}{2} (js - kr)} \, e_{j+r,k+s} 
$$
from which:
\beq\label{commutrel-rt}
[e_{j,k}, e_{r,s}] = 2 \ii \sin \frac{\pi}{N} (js - kr) \, e_{j+r,k+s}
\eeq
Also, 
\begin{align*}
e_{r,s} & = \omega^{\frac{1}{2} rs}  Q^{-s} P^{-r} = \omega^{\frac{1}{2} rs} 
\sum_{m=0}^{N-1} \omega^{-ms} \ket{m}\bra{m} \, P^{-r}\\
& = e_{-r,-s} \end{align*}
and
\begin{align}
e_{N \pm r, s} & = (-1)^{s} \, e_{\pm r, s} , \qquad  e_{r,N \pm s} = (-1)^{r} \, e_{r, \pm s} \quad \Rightarrow \label{idU} \\
& e_{N+r,N+s}  = (-1)^{N + r+s} \, e_{r,s} , \nn \\
& e_{N-r,N-s} = (-1)^{N - r - s} \, e_{-r,-s} = (-1)^{N - r - s} \, e^*_{r,s} 
\nn \end{align}
with their conjugated. 

\noindent
We define `truncated' matrices for $0 < a < N $ and $b\in \Z$, as follow
\beq\label{eff}
f_{a,b} 
= \omega^{- \frac{1}{2} ab} R^a Q^b
= \omega^{- \frac{1}{2} ab} P^a \sum_{n=a}^{N-1} \omega^{bn} \ket{n} \bra{n}
\eeq
\beq\label{efft}
\wt{f}_{a,b} 
= \omega^{\frac{1}{2} ab}  Q^{-b} R^{N-a}
= \omega^{\frac{1}{2} ab}  \sum_{n=0}^{a-1} \omega^{-bn} \ket{n} \bra{n} P^{(N-a)} 
\eeq
We may indeed define them also for $a=0$ and $a=N$ observing from the above that 
\beq\label{effts0N}
f_{0,b} = e_{0,b}, \quad f_{N,b} = 0 , \qquad \quad 
\wt{f}_{0,b} = 0 , \quad \wt{f}_{N,b} = (-1)^b e^*_{0,b}. 
\eeq
When $a\not= 0, \not= N$, $f_{a,b}$ is the upper triangular part of $e_{a,b}$ while $\wt{f}_{a,b}$
is the upper triangular part of $e^*_{a,b}$.
One finds, for $0 \leq a \leq N-1$ with $0 \leq N-a \leq N-1$ and $b\in \Z$,
\begin{align}  \label{idT}
f_{0, -b} & = \wt{f}_{0,b} \qquad \wt{f}_{0, -b} = f_{0,b} \\
f_{a,N \pm b} & =  (-1)^{- a} f_{a, \pm b}  \qquad \wt{f}_{a,N \pm b} =  (-1)^{a} \wt{f}_{a, \pm b} \nn \\ 
f_{N-a,b} & = (-1)^{-b} \wt{f}_{a,-b}   \qquad \wt{f}_{N-a,b} = (-1)^{b} f_{a,-b}  \nn \\ 
\Rightarrow \quad \wt{f}_{N-a,N-b} & =  (-1)^{N + a - b} f_{a,b} \nn \\ 
f_{N-a,N-b}  & =  (-1)^{-N - a + b} \wt{f}_{a,b}. 
\end{align}
Moreover, 
\begin{align}
f_{a,b} f_{r,s} & = \omega^{\frac{1}{2} ( a s - b r ) } f_{a+r,b+s} \qquad = 0 \quad \mbox{if} \quad a+r \geq N \nn \\
\wt{f}_{a,b} \wt{f}_{r,s} & = (-1)^{b+s} \omega^{\frac{1}{2} ( a s - b r ) } 
\wt{f}_{a+r -N,b+s} \qquad = 0 \quad \mbox{if} \quad a+r \leq N
\end{align}
The above imply:
$$
[f_{j,k}, f_{a,b}] = 2 \ii  \sin \frac{\pi}{N} (jb - ka) \, f_{j+a,k+b} \qquad = 0 \quad \mbox{if} \quad j + a \geq N
$$
\begin{align}
[f_{j,k}, \wt{f}_{a,b}] &= - 2 \ii \sin \frac{\pi}{N} (jb - ka) \, \wt{f}_{a-j,b-k} \nn \\
 & =  - (-1)^{k+b} \, 2 \ii \sin \frac{\pi}{N} (jb - ka) \, f_{N+j-a, k-b} \qquad = 0 \quad \mbox{if} \quad a - j \leq 0 
\end{align}
$$
[\wt{f}_{j,k}, \wt{f}_{a,b}] = (-1)^{k+b} \, 2 \ii \sin \frac{\pi}{N} (jb - ka) \, \wt{f}_{j+a-N,k+b} 
\qquad = 0 \quad \mbox{if} \quad j + a \leq N .
$$

\subsection{The Cartan sub-algebra}
The sub-algebra $\ft$ of diagonal matrices with purely imaginary entries is given as: 
\begin{align}
U_{0,b} & = \frac{\ii}{\sqrt{N}} \, ( e_{0,b} + e^*_{0,b} ) = 
\frac{2\ii}{\sqrt{N}} \sum_{m=0}^{N-1} \cos (2 \pi \frac{mb}{N}) \ket{m} \bra{m} \\
\wt{U}_{0,b} & = \frac{1}{\sqrt{N}} ( e_{0,b} - e^*_{0,b} ) = \frac{2 \ii}{\sqrt{N}} \sum_{m=0}^{N-1} \sin (2 \pi \frac{mb}{N}) \ket{m} \bra{m}
\end{align}
for $b = 0,1, \cdots N-1$. Being $U_{0,N-b} = - U_{0,b}$ and $\wt{U}_{0,N-b}=\wt{U}_{0,b}$, a basis of $\ft$ is then given by
\beq\label{bcar}
( \ii H_b, \ii H_{\tilde b} ) = 
\begin{cases}
\{U_{0,b}, \, b = 0,1, \cdots L \}  \bigcup \{ \wt{U}_{0,b} , \, b = 1, \cdots, L\}  \qquad N = 2L + 1\\
~\\
\{U_{0,b}, \, b = 0,1, \cdots L-1 \}  \bigcup \{ \wt{U}_{0,b} , \, b = 1, \cdots, L\}  \qquad N = 2L 

\end{cases}
\eeq
with cardinality $N$ in both cases. One has $H_L = 0$ when $N=2L$.

\noindent
Now $\ft$ is the maximal abelian sub-algebra of $\u{U}_N$, the skew symmetric matrices, and $\fh = \ft \oplus  \ii \ft$ is the Cartan sub-algebra of $\u{GL}_N$. 

\subsection{Upper-triangular matrices and Borel sub-algebra}
Due to \eqref{effts0N}, we could also write
\beq\label{bcarup}
H_b = \frac{1}{\sqrt{N}} \, (f_{0,b} + \wt{f}_{N,b}) \qquad 
H_{\tilde b} = - \frac{\ii}{\sqrt{N}} \, (f_{0,b} - \wt{f}_{N,b}) .
\eeq
These diagonal matrices have real entries and are `diagonal upper triangular' matrices.
The sub-algebra $\fn_+$ of strictly upper-triangular matrices is made of the matrices,
\beq\label{sut}
T^{a,b} = \frac{1}{\sqrt{N}} (f_{a,b} + \wt{f}_{a,b}) \qquad 
\wt{T}^{a,b} = - \frac{\ii}{\sqrt{N}} (f_{a,b} - \wt{f}_{a,b}) \qquad  0 < a < N . 
\eeq
Owing to relations \eqref{idT} not all matrices in \eqref{sut} are independent (some of them may indeed vanish).  
A basis for the strict upper triangular ones is given by
\beq\label{buup}
 \big\{T^{r,s} , \, \wt{T}^{r,s} , \, \, (r,s) \in \{1, \cdots, L \} \times \{0, 1, \cdots, N-1 \}  \big\} 
 \qquad N = 2L +1 
 \eeq
\begin{multline}\label{buup1}
 \big\{T^{r,s} , \, \wt{T}^{r,s} , \, \, (r,s) \in \{1, \cdots, L - 1\} \times \{0, 1,\cdots, N-1 \}  \big\} \\
 \bigcup
 \{T^{L,s} , \, \, s = 1, \cdots, L-1 \}   \bigcup  \{\wt{T}^{L,s} , \, \, s = 0, \cdots, L \} \qquad N = 2L .
\end{multline}
Note that both $T^{L,0}$ and $T^{L,L}$ vanish when $N=2L$. The basis is 
of cardinality $N(N-1)$ in both cases.

The (upper) Borel sub-algebra is then $\u{B}_N =  \ii \ft \oplus \fn_+$, with basis the union 
of \eqref{buup}, \eqref{buup1}, and real matrices $(H_b, H_{\tilde b})$ as in \eqref{bcarup} adding up to $N^2$ elements. 

\subsection{The anti-hermitian matrices}
The sub-algebra $u_N$ of anti-hermitian matrices: 
\beq\label{ahm}
U_{r,s} = \frac{\ii}{\sqrt{N}}\, (e_{r,s} + e^*_{r,s}) \qquad 
\wt{U}_{r,s} = \frac{1}{\sqrt{N}} \, (e_{r,s} - e^*_{r,s})
\eeq
In particular we have
$$
U_{0,s} = \frac{\ii}{\sqrt{N}}\, (e_{0,s} + e^*_{0,s}) = \ii H_s \qquad \qquad 
\wt{U}_{0,s} = \frac{1}{\sqrt{N}} \, (e_{0,s} - e^*_{0,s}) = \ii H_{\tilde s}  
$$
making up the maximal abelian $N$-dimensional sub-algebra $\ft$ of $u_N$.

Again, due to relations \eqref{idU} of the matrices in \eqref{ahm} the independent ones are $N(N-1)$ in number.
A basis for the non-diagonal ones is given by
\beq\label{bah}
 \big\{U_{r,s} , \, \wt{U}_{r,s} , \, \, (r,s) \in \{1, \cdots, L \} \times \{0, 1, \cdots, N-1 \}  \big\} 
 \qquad N = 2L +1 
 \eeq
\begin{multline}\label{bah1}
 \big\{U_{r,s} , \, \wt{U}_{r,s} , \, \, (r,s) \in \{1, \cdots, L - 1\} \times \{0, 1,\cdots, N-1 \}  \big\} \\
 \bigcup
 \{U_{L,s} , \, \, s = 1, \cdots, L-1 \}   \bigcup  \{\wt{U}_{L,s} , \, \, s = 0, \cdots, L \} \qquad N = 2L .
\end{multline}
Now both $U_{L,0}$ and $U_{L,L}$ vanish when $N=2L$. The basis is 
of cardinality $N(N-1)$ in both cases, adding to $N^2$, the dimension of $\u{U}_N$, 
with the $N$ diagonal ones in \eqref{bcar}.

\subsection{The Lie bi-algebra structure}
From the above it follows that as real vector spaces
\beq\label{sub}
\u{GL}_N = \u{U}_N \oplus \u{B}_N
\eeq
\begin{lemma}
The subspaces $\u{U}_N$ and $\u{B}_N$ are isotropic and dual via the invariant pairing:
\beq\label{IP}
\hs{X}{Y} = \textup{Im} \, \tr (X  Y) 
\eeq
the imaginary part of the trace of the product. 
\end{lemma}
\noindent
The invariance is just: $\hs{[Z,X]}{Y} + \hs{X}{[Z,Y]} = 0$. 
The explicit proof is in Appendix \ref{app1} where it is shown that both $u_N$ and $\fb_N$ are isotropic and that 
the only non vanishing pairings are 
\beq\label{parnot00}
\hs{T^{a,b}}{U_{a, b}} = 1 = \hs{\wt{T}^{a,b}}{\wt{U}_{a, b}}  \qquad a,b = 0,1, \cdots , N-1.
\eeq
Thus, the basis \eqref{buup}, \eqref{buup1} for the strictly upper triangular matrices of $\fb_N$ is dual to the basis \eqref{bah}, \eqref{bah1} of anti-hermitian non diagonal matrices, while the basis $(H_b, H_{\tilde b})$ in \eqref{bcar}
for the real diagonal matrices is dual to the basis $(\ii H_b, \ii H_{\tilde b})$ for the diagonal purely imaginary matrices. 

The data $(\u{GL}_N, \u{U}_N, \u{B}_N)$ makes up a Manin triple and then \cite[Prop.~1.3.4]{ChariPressley} a Lie bi-algebra. 

\section{The classical limit $N\to \infty$}\label{sec:ClassicalLimit}
Consider the large $N$ limit of the construction of the previous section. 
As for the commutation relations \eqref{commutrel-rt}, for large $N$ with $j,k,r,s$ fixed, one gets
$$
[e_{j,k}, e_{r,s}] \sim \ii \frac{2 \pi}{N} (js - kr) \, e_{j+r,k+s}
$$
This goes to the usual Poisson structure on the torus 
\begin{align}\label{clalim}
\{e_{j,k}, e_{r,s}\} & \sim \frac{\ii}{\hbar} [e_{j,k}, e_{r,s}] \, _{\big |_{\hbar = 0}} = - (js - kr) \, e_{j+r,k+s}
\end{align}
and
$\frac{2 \pi}{N}$ is the analogue of $\hbar$. 
Also, the limit of anti-hermitian matrices $U_{N=\infty}$ is the Lie algebra of area preserving vector fields on the commutative torus. We shall next describe the Lie bi-algebra structure on the (commutative) algebra $C^\infty(\T^2)$ obtained in the limit.

\subsection{The algebra} 
The (commutative) algebra $C^\infty(\T^2)$ 
of complex valued smooth functions on the torus $\T^2$ is made of all elements
of the form 
\beq\label{2ta}
\phi = \sum_{(m_1,m_2) \in \Z^2} \phi_{m_1 m_2} \, e^{\ii m_1 x_1 + m_2 x_2} 
= \sum_{m \in \Z^2} \phi_{m} \, e^{\ii m \cdot x} 
\eeq
with coefficients $\{\phi_{mn}\}\in S(\Z^2)$ a complex-valued Schwartz function
on $\Z^2$. This means that the sequence of complex numbers
$\{\phi_{m,n} \in \C~|~ (m,n) \in\Z^2 \}$ decreases rapidly at
`infinity', that is  for any $k = 0, 1, 2, \cdots $, one has bounded semi-norms
\beq\label{nctsn}
|| \phi ||_k = \sup_{(m,n)\in \Z^2} ~|\phi_{m,r}|\,\big(1+ |m|+|r| \big)^k < \infty ~.
\eeq
The algebra $C^\infty(\T^2)$ is a Poisson algebra with brackets:
$$
\{\phi, \psi\} = \partial_1\phi \partial_2 \psi - \partial_2 \phi \partial_1 \psi
$$
For the basis elements $e_m = e^{\ii m \cdot x}$ one gets:
\beq\label{cpb}
\{e_k, e_m\} = - k \wedge m \, e_{k+m} \qquad k \wedge m = k_1 m_2 - k_2 m_1
\eeq

An invariant real valued inner product, the counterpart of \eqref{IP}, is given by 
\beq\label{ripc}
\hs{\phi}{\psi} = \textup{Im} \, \int \phi \psi \, \frac{d^2 x}{(2 \pi)^2} = \frac{1}{2 \ii} \, \sum_{m\in\Z^2} (\phi_m \psi_{-m} - \phi^*_{-m} \psi^*_{m} ). 
\eeq
The sum is finite due to the condition \eqref{nctsn} on the coefficients. The invariance means that 
$$
\hs{ \{\eta, \phi \} }{\psi} + \hs{ \phi }{ \{\eta, \psi \} } = 0 .
$$
The inner product is non degenerate but is not positive definite. 
We use the inner product to break the Poisson Lie algebra $C^\infty(\T^2)$, regarded as a {\it real} Lie algebra, as the sum of two real sub-algebras which are isotropic and paired via the inner product. We seek a splitting of the kind in \eqref{sub} of a `real' and `upper triangular' parts (rather then a `purely imaginary' part).

Start with the real Lie sub-algebra of real functions:
$$
\u{A} = \Big\{ \phi = \sum\nolimits_{m \in \Z^2} \phi_{m} \, e_m \, | \, \phi^*_{m} = \phi_{-m} \Big\} 
$$
This is isotropic:
$$
\hs{\phi}{\psi} = \frac{1}{2 \ii} \, \sum_{m\in\Z^2} (\phi_m \psi_{-m} - \phi_{m} \psi_{-m} ) = 0.  
$$
It is convenient to use the trigonometric basis for $\u{A}$. 
To avoid over-counting indices are restricted, following a lexicographic ordering:
\beq\label{basa1}
U_m =
\begin{cases}
\, \tfrac{1}{2} (e_m + e^*_m) = \cos (m \cdot x) \qquad & m_1 > 0 \, , m_2 \in \Z \\ 
\, \tfrac{1}{2} (e_{0,m_2} + e^*_{0,m_2}) = \cos (m_2 x_2) \qquad & m_2 > 0 \\
\, 1 \qquad & m = 0
\end{cases} 
\eeq 
\beq\label{basa2}
\wt{U}_m =
\begin{cases}
\, - \tfrac{\ii}{2} (e_m - e^*_m) = \sin (m \cdot x) \qquad & m_1 > 0 \, , m_2 \in \Z \\ 
\, - \tfrac{\ii}{2} (e_{0,m_2} - e^*_{0,m_2}) = \sin (m_2 x_2) \qquad & m_2 > 0 
\end{cases} 
\eeq 

\noindent
Next, consider the real Lie sub-algebra of functions:
$$
\u{B} = \Big\{ \psi = \sum\nolimits_{m_1 \geq 0, m_2 \in \Z} \psi_{m} \, e_m \, | \, \psi^*_{0,m_2} =  -\psi_{0,-m_2} \Big\} 
$$
So, $\u{B}$ is made of functions on the torus that can be continued 
holomorphically to the disk in the first variable, and are purely imaginary when averaged over $x_1$:
$$
\psi = \sum\nolimits_{m_1 \geq 0, m_2 \in \Z} \psi_{m} \, z_1^{m_1} \, e^{\ii m_2 x_2} , \quad z_1 
= e^{\ii x_1} , \quad |z_1| < 1. 
$$
The sub-algebra $\u{B}$ is isotropic as well:
$$
\hs{\phi}{\psi} = \frac{1}{2 \ii} \, \sum_{n \in\Z} (\phi_{0,n} \psi_{0,-n} - \phi_{0,n} \psi_{0,-n} ) = 0.  
$$
A basis for $\u{B}$ is given by: 
\beq\label{basb1}
T^n =
\begin{cases}
\, 2 \ii e_n \qquad & n_1 > 0 \, , n_2 \in \Z \\ 
\, \ii (e_{0,n_2} + e^*_{0,n_2}) = 2 \ii \cos (n_2 x_2) \qquad & n_2 > 0  \\
\, \ii  \qquad & n=0
\end{cases} 
\eeq 
\beq\label{basb2}
\wt{T}^n =
\begin{cases}
\, 2 e_n \qquad & n_1 > 0 \, , n_2 \in \Z \\ 
\, e_{0,n_2} - e^*_{0,n_2} = 2 \ii \sin (n_2 x_2) \qquad & n_2 \geq 0 
\end{cases} 
\eeq 

\noindent
The basis \eqref{basa1} and \eqref{basa2} is dual to the basis \eqref{basb1} and \eqref{basb2} for the inner product \eqref{ripc} as it can be checked directly.  
The only non vanishing pairings are:
\beq\label{parfuncnct}
\hs{T^m}{U_m} = 1 = \hs{\wt{T}^m}{\wt{U}_m} .
\eeq

\subsection{The commutation relations}
Out of \eqref{cpb}, we workout explicitly the commutation relations of the generators $U$'s of the algebra $A$ in 
\eqref{basa1} and \eqref{basa2}, and of the generators $T$'s of the algebra $B$ in \eqref{basb1} and \eqref{basb2}. 
Let $m=(m_1, m_2)$, $n=(n_1, n_2)$ be two elements in $\Z^2$. 
\noindent
\subsubsection{The algebra $\u{A}$}
Firstly: $\{U_m, U_n \}=\{U_m, \wt{U}_n \}= \{\wt{U}_m, \wt{U}_n \} =0$, when $m_1= n_1 = 0$.

\noindent
Then,
\begin{align}\label{UU}
m_1 > 0, \, n_1 > 0 \qquad \{U_m, U_n \} 
& = \tfrac{1}{2} m \wedge n \, (- U_{m+n} + U_{m-n}) \qquad \mbox{if} \quad m_1 > n_1 \nn \\
& = \tfrac{1}{2} m \wedge n \, (- U_{m+n} + U_{n-m}) \qquad \mbox{if} \quad m_1 < n_1 
\end{align}
\begin{align}\label{U0U}
m_1= n_1 > 0 \qquad \{U_m, U_n \} 
& = \tfrac{1}{2} m_1 (n_2 - m_2) \, (- U_{m+n} + U_{0, m_2-n_2}) \qquad \mbox{if} \quad m_2 > n_2 \nn \\
& = \tfrac{1}{2} m_1 (n_2 - m_2) \, (- U_{m+n} + U_{0, n_2-m_2}) \qquad \mbox{if} \quad m_2 < n_2 .
\end{align}
Relations \eqref{UU} is valid also when either $m_1 = 0$ or $n_1 = 0$, but not both. 

\noindent
Next, 
\begin{align}\label{UUt}
m_1 > 0, \, n_1 > 0 \qquad \{U_m, \wt{U}_n \} 
& = - \tfrac{1}{2} m \wedge n \, (\wt{U}_{m+n} + \wt{U}_{m-n}) \qquad \mbox{if} \quad m_1 > n_1 \nn \\
& = - \tfrac{1}{2} m \wedge n \, (\wt{U}_{m+n} - \wt{U}_{n-m}) \qquad \mbox{if} \quad m_1 < n_1 
\end{align}
\begin{align}\label{U0Ut}
m_1= n_1 > 0 \qquad \{U_m, \wt{U}_n \} 
& = - \tfrac{1}{2} m_1 (n_2 - m_2) \, (\wt{U}_{m+n} + \wt{U}_{0, m_2-n_2}) \qquad \mbox{if} \quad m_2 > n_2 \nn \\
& = - \tfrac{1}{2} m_1 (n_2 - m_2) \, (\wt{U}_{m+n} - \wt{U}_{0, n_2-m_2}) \qquad \mbox{if} \quad m_2 < n_2 .
\end{align}
Relations \eqref{UUt} is valid also when either $m_1 = 0$ or $n_1 = 0$, but not both. 

\noindent
Finally, 
\begin{align}\label{UtUt}
m_1 > 0, \, n_1 > 0 \qquad \{\wt{U}_m, \wt{U}_n \} 
& = \tfrac{1}{2} m \wedge n \, (U_{m+n} + U_{m-n}) \qquad \mbox{if} \quad m_1 > n_1 \nn \\
& = \tfrac{1}{2} m \wedge n \, (U_{m+n} + U_{n-m}) \qquad \mbox{if} \quad m_1 < n_1 
\end{align}
\begin{align}\label{Ut0Ut}
m_1= n_1 > 0 \qquad \{\wt{U}_m, \wt{U}_n \} 
& = \tfrac{1}{2} m_1 (n_2 - m_2) \, (U_{m+n} + U_{0, m_2-n_2}) \qquad \mbox{if} \quad m_2 > n_2 \nn \\
& = \tfrac{1}{2} m_1 (n_2 - m_2) \, (U_{m+n} + U_{0, n_2-m_2}) \qquad \mbox{if} \quad m_2 < n_2 .
\end{align}
Relations \eqref{UtUt} is valid also when either $m_1 = 0$ or $n_1 = 0$, but not both. 

\noindent
\subsubsection{The algebra $\u{B}$}
Firstly: $\{T^m, T^n \}=\{T^m, \wt{T}^n \}=\{\wt{T}^m, \wt{T}^n \}=0$ when $m_1= n_1 = 0$.

\noindent
Then,
\begin{align}
m_1 > 0, \, n_1 > 0 \qquad \{T^m, T^n \} & = 2 m \wedge n \, \wt{T}^{m+n}  \label{TT} \\
m_2 > 0, \, n_1 > 0 \qquad \{T^{(0,m_2)}, T^n \} & = - m_2 n_1 \, (\wt{T}^{(n_1, n_2+m_2)}  
- \wt{T}^{(n_1, n_2-m_2)}) . \label{T0T}
\end{align}
Next, 
\begin{align}
m_1 > 0, \, n_1 > 0 \qquad \{T^m, \wt{T}^n \} & = - 2 m \wedge n \, T^{m+n}  \label{TTt} \\
m_2 > 0, \, n_1 > 0 \qquad \{T^{(0,m_2)}, \wt{T}^n \} & = m_2 n_1 \, (T^{(n_1, n_2+m_2)} - T^{(n_1, n_2-m_2)}) . \label{T0Tt} \\
m_2 > 0, \, n_1 > 0 \qquad \{\wt{T}^{(0,m_2)}, T^n \} & = m_2 n_1 \, (T^{(n_1, n_2+m_2)} + T^{(n_1, n_2-m_2)}) . \label{Tt0T}
\end{align}
Finally, 
\begin{align}
m_1 > 0, \, n_1 > 0 \qquad \{\wt{T}^m, \wt{T}^n \} & = - 2 m \wedge n \, \wt{T}^{m+n}  \label{TtTt} \\
m_2 > 0, \, n_1 > 0 \qquad \{\wt{T}^{(0,m_2)}, \wt{T}^n \} & = m_2 n_1 \, (\wt{T}^{(n_1, n_2+m_2)}  
+ \wt{T}^{(n_1, n_2-m_2)}) . \label{Tt0Tt}
\end{align}

\subsection{The Lie bi-algebra structure} 
We indicate by $\Gamma$ the structure constants of the subalgebra $\u{A}$ and by $\Delta$ those of the subalgebra 
$\u{B}$. The mixed commutators ought to be of the form 
$$
\{T^a, U_b \} = \Gamma^a_{bd} T^d - \Delta_b^{ad} U_d .
$$
Let $m=(m_1, m_2)$, $n=(n_1, n_2)$ be two elements in $\Z^2$.

$\bullet$ \, $m_1>0 , n_1>0, m_1 \not= n_1$:
$$
\{T^m, U_n\} = - \ii \, m \wedge n (e_{m+n} - e_{m-n})
$$

\noindent
$m_1 < n_1$: the admissible structure constants are
$$
\Delta_n^{m, \wt{n-m}} , \quad \Gamma^m_{n, n+m} , \quad \Gamma^m_{n, n-m}
$$
then
$$
\Gamma^m_{n, n+m} T^{n+m} + \Gamma^m_{n, n-m} T^{n-m} - \Delta_n^{m, \wt{n-m}} \wt{U}_{n-m} 
= - \ii \, m \wedge n (e_{m+n} - e_{m-n})
$$

\noindent
$m_1 > n_1$: the admissible structure constants are
$$
\Gamma^m_{n, m+n} , \quad \Gamma^m_{n, m-n}
$$
then
$$
\Gamma^m_{n, m+n} T^{m+n} + \Gamma^m_{n, m-n} T^{m-n} = - \ii \, m \wedge n (e_{m+n} - e_{m-n})
$$

\medskip
$\bullet$ \, $m_1 = n_1>0$:
$$
\{T^m, U_n\} = \ii \, m_1(m_2 - n_2) (e_{m+n} - e_{0, m_2-n_2})
$$

\noindent
$m_2 < n_2$: the admissible structure constants are
$$
\Delta_n^{m, (0, \wt{n_2-m_2})} , \quad \Gamma^m_{n, n+m} , \quad \Gamma^m_{n, (0, n_2-m_2)}
$$
then
$$
\Gamma^m_{n, n+m} T^{n+m} + \Gamma^m_{(0, n_2-m_2)} T^{(0, n_2-m_2)} - \Delta_n^{m, (0, \wt{n_2-m_2})} 
\wt{U}_{(0, n_2-m_2)} 
= \ii \, m_1(m_2 - n_2) (e_{m+n} - e_{0, m_2-n_2})
$$

\noindent
$m_2 > n_2$: the admissible structure constants are
$$
\Delta_n^{m, (0, \wt{m_2-n_2})} , \quad \Gamma^m_{n, m+n} , \quad \Gamma^m_{n, (0, m_2-n_2)}
$$
then
$$
\Gamma^m_{n, m+n} T^{m+n} + \Gamma^m_{(0, m_2-n_2)} T^{(0, m_2-n_2)} - \Delta_n^{m, (0, \wt{m_2-n_2})} 
\wt{U}_{(0, m_2-n_2)} 
= \ii \, m_1(m_2 - n_2) (e_{m+n} - e_{0, m_2-n_2})
$$
 
\medskip 
$\bullet$ \, $m_1=0 , n_1>0$:
$$
\{T^{(0,m_2)}, U_n\} = \tfrac{\ii}{2} \, m_2 n_1 (e_{(n_1, m_2+n_2)} + e_{(-n_1, -m_2-n_2)} 
- e_{(n_1, - m_2+n_2)} - e_{(-n_1, m_2-n_2)})
$$

\noindent
the admissible structure constants are
$$
\Delta_n^{(0,m_2), \wt{(n_1, n_2+m_2})} , \quad \Delta_n^{(0,m_2), \wt{(n_1, n_2-m_2})} , 
\quad \Gamma^{(0,m_2)}_{n, (n_1, n_2+m_2)} , \quad \Gamma^{(0,m_2)}_{n, (n_1, n_2-m_2)}
$$
then
\begin{align*}
\Gamma^{(0,m_2)}_{n, (n_1, n_2+m_2)} T^{n_1, n_2+m_2} &+ \Gamma^{(0,m_2)}_{n, (n_1, n_2-m_2)} T^{n_1, n_2-m_2} \\ 
& \qquad - \Delta_n^{(0,m_2), \wt{(n_1, n_2+m_2})} \wt{U}_{n_1, n_2+m_2} 
- \Delta_n^{(0,m_2), \wt{(n_1, n_2-m_2})} \wt{U}_{n_1, n_2-m_2} \\
& = \tfrac{\ii}{2} \, m_2 n_1 (e_{(n_1, m_2+n_2)} + e_{(-n_1, -m_2-n_2)} 
- e_{(n_1, - m_2+n_2)} - e_{(-n_1, m_2-n_2)})
\end{align*}

\medskip
$\bullet$ \, $m_1 >0 , n_1=0$:
$$
\{T^m, U_n\} = - \ii \, m_1 n_2 (e_{(m_1, m_2+n_2)} - e_{m_1, m_2-n_2})
$$

\noindent
the admissible structure constants are
$$
\Gamma^m_{(0,n_2), (m_1, m_2+n_2)} , \quad \Gamma^m_{(0,n_2), (m_1, m_2-n_2)}
$$
then
$$
\Gamma^m_{(0,n_2), (m_1, m_2+n_2)} T^{m_1, m_2+n_2} + \Gamma^m_{(0,n_2), (m_1, m_2-n_2)} T^{m_1, m_2-n_2} 
= - \ii \, m_1 n_2 (e_{(m_1, m_2+n_2)} - e_{m_1, m_2-n_2})
$$

All  other commutators,  $\{T^a, \wt{U}_b \}$, $\{\wt{T}^a, U_b \}$, $\{\wt{T}^a, \wt{U}_b \}$,
go along the same lines.

\section{The  non-commutative torus}
\subsection{The algebra}
Let $\theta$ be a real number. The algebra $A_\theta =C^\infty(\T^2_\theta)$ 
of smooth functions on the
non-commutative torus $\T^2_\theta$ is the associative algebra made up of all elements
of the form, 
\beq\label{2tanc}
a =\sum_{(m,n) \in \Z^2} a_{mn} \, Q^m P^n ,
\eeq
with two generators $Q$ and $P$ that satisfy 
\beq\label{nct}
P\,Q =e^{2\pi \ii  \theta} \, Q\,P .
\eeq
As in the commutative case, the coefficients $\{a_{mn}\}\in S(\Z^2)$ form a complex-valued Schwartz function
on $\Z^2$. The algebra $A_\theta$ can be made into a $*$-algebra by
defining an involution  by
\beq\label{nctsta}
Q^\dagger := Q^{-1} \, , ~~~ P^\dagger := P^{-1}\, ,
\eeq
so that $Q$ and $P$ are unitary.
Heuristically, the non-commutative relation \eqref{nct} of the torus
is the exponential of the Heisenberg commutation relation
$[x_2,x_1]= \ii \theta/2\pi$. The algebra $A_\theta$  can be represented 
as bounded operators on the Hilbert space $H=L^2(\R)$ by
\beq\label{hhs}
(Q f)(t) = e^{2\pi \ii t} \, f(t) \qquad (P f)(t) = f(t+\theta) .
\eeq
getting the commutation relations \eqref{nct}. 

From \eqref{nct} one sees that $A_\theta$ is commutative if
and only if $\theta$ is an integer, and one identifies $ A_0$
with the algebra $C^\infty(\T^2)$ of complex-valued smooth functions on
an ordinary square two-torus $\T^2$ with coordinate
functions given by $Q =e^{\ii x_1}$ and $P =e^{\ii x_2}$, recovering then the 
Fourier expansion \eqref{2ta} of any such a function. 

When the deformation parameter is a rational number, $\theta=M/N$, with $M$ and $N$
positive integers (taken to be relatively prime, say) also the algebra
$ A_{M/N}$ is 
related to the algebra $C^\infty(\T^2)$. 
More precisely, $A_{M/N}$ is Morita equivalent to
$C^\infty(\T^2)$, that is  $ A_{M/N}$ is a twisted matrix bundle over
$C^\infty(\T^2)$ of Chern number $M$ whose fibers are $N\times N$
complex matrix algebras.
The algebra $ A_{M/N}$ has a
`huge' center $C( A_{M/N})$ which is generated by the elements
$Q^N$ and $P^N$. One identifies $C( A_{M/N})$ with the algebra
$C^\infty(\T^2)$ = of the torus winding $N$ times over itself, 
while there are finite dimensional representations given as copies of those in \eqref{csm}.
 
 Denote $\omega =e^{2 \pi \ii \theta}$. It is convenient to change basis to
 $$
\hat{e}_m = \omega^{- \frac{1}{2} m_1 m_2} P^{m_1} Q^{m_2} \qquad m = (m_1, m_2) \in \Z .
 $$
Then $\hat{e}^\dagger{}_m=\hat{e}_{-m}$. From \eqref{hhs} they act on $H=L^2(\R)$ as 
\beq\label{hhs1}
(\hat{e}_m f)(t) = \omega^{\frac{1}{2} m_1 m_2} \, e^{2\pi \ii \, m_2 t} \, f(t+m_1 \theta) .
\eeq
These elements yield an infinite-dimensional Lie algebra (the sine-algebra). Indeed, one checks
$$
\hat{e}_k \hat{e}_m = \omega^{\frac{1}{2} k \wedge m} \, \hat{e}_{k+m} \qquad k \wedge m = k_1 m_2 - k_2 m_1
$$ 
 This implies 
\beq\label{cnct}
\hat{e}_k \hat{e}_m - \hat{e}_m \hat{e}_k = 2 \ii \sin ( \pi \theta \, k \wedge m) \, \hat{e}_{k+m} .
\eeq
The sine-algebra has the role of the hamiltonian vector fields on $\T^2$ for the canonical Poisson structure.
The above is indeed seen as the quantisation of the canonical Poisson structure on $\T^2$. As a vector space 
$A_\theta$ and $C^\infty(\T^2)$ are the same. With an abuse of notation on generators, $\hat{e}_m \to e_m$, the Poisson structure is recovered as
\begin{align}
\{e_k, e_m\} & = \frac{\ii}{\hbar} (e_k e_m - e_m e_k) \, _{\big |_{\hbar \sim \theta = 0}} \nn \\
& = - \frac{2 \pi}{\hbar} \theta \, k \wedge m \, e_{k+m}  \label{clalim-nc} \\
& = - k \wedge m \, e_{k+m}  \nn
\end{align}
with the identification $\hbar = 2 \pi \theta$, in parallel with \eqref{clalim}. This is just the Poisson structure \eqref{cpb}.

\subsection{The trace and the $K$-theory}
On the algebra $ A_\theta$ there is a (unique if $\theta$ is irrational) normalized, positive definite trace, 
$\tau :  A_\theta \to \C$, given by
\beq\label{ncintadefa00}
\tau ( \sum_{m \in \Z^2} a_m \, \hat{e}_m ) := a_{0}.
\eeq
Then, for any $a,b\in A_\theta$, one checks that 
\beq
\tau(a\,b) = \sum_{m \in \Z^2} a_m \, b_{-m} = \tau(b\,a).
 \eeq
Also, $\tau(1) = 1$, $\tau(a^\dag\,a)  > 0$, for $a\not= 0$ 
and $\tau(a^\dag a) = 0$ if and only if $a=0$ (the trace is faithful).

This trace is invariant under the natural action of the
commutative torus $\T^2$ on $\A_\theta$ whose infinitesimal form
is generated by two commuting derivations $\partial_1, \partial_2$ acting
as
\beq\label{t2act}
\partial_1 (P) = 2\pi \ii P , \quad \partial_1 (Q) = 0 , \qquad \partial_1 (P) = 0 , \quad \partial_1 (Q) = 2\pi \ii Q
\eeq
Invariance is just the statement that $\tau( \partial_1 (a)) = 0 = \tau( \partial_1 (a)) $ for $a\in A_\theta$.

A remarkable fact about the non-commutative two-torus algebra $A_\theta$ is that it contains not trivial projections. 
In fact it contains a representative projection for each equivalence classe in the $K$-theory of the torus. 
The archetype of all such projections is the Powers-Rieffel projection \cite{Ri81}. 
To construct it, observe first that
there is an injective algebra homomorphism
\begin{align}
\rho \, : \,C^\infty\left(\IS^1\right) & \longrightarrow A_\theta \, , \nn \\
f(x_1) \, = \sum_{m\in\Z}f_{m} \, e^{2 \pi \ii m\,x_1} & \longmapsto \rho(f) = \sum_{m\in\Z}f_{m} \, Q^m .
\label{xsubalg}
\end{align}
From the commutation relations \eqref{nct} it follows that if $f(x_1)$ is mapped to $\rho(f)$, then 
$P \rho(f) P^{-1}$ is the image of the shifted function $f(x_1+\theta)$.
One now looks for projections of the form
\beq
p_\theta = P^{-1} \rho(g)+\rho(f)+\rho(g) P \, .
\label{Ptheta}
\eeq
In order that
\eqref{Ptheta} defines a projection operator $p^2=p$, the functions $f,g \in
C^\infty(\IS^1)$ must satisfy some conditions.  
These conditions are satisfied by the choice 
\begin{eqnarray}\label{fgproj}
&& f(x_1) = \left\{
\begin{array}{ccrcl}
\mbox{\rm smoothly increasing from 0 to 1} & ~ & 0 & \leq~x_1~\leq & 1-\theta \\~\\
1 & ~ &  1 - \theta & \leq~x_1~\leq &  \theta \\ ~\\
1 - f(x_1-\theta) & ~ & \theta & \leq~x_1~\leq &  1 \\
\end{array}
\right. \ , \nn \\ &&{~~~~}^{~~}_{~~} \nn \\
&& g(x_1) = \left\{
\begin{array}{ccrcl}
0 & ~ & 0 &\leq~x_1~\leq & \theta  \\ & \\
\sqrt{f(x_1)-f(x_1)^2} & ~ & \theta & \leq~x_1~\leq & 1
\end{array}\right. \ .
\label{bumptheta}
\end{eqnarray}

\noindent
It is straighforward to check that the rank (i.e. the trace) of
$p_\theta$ is just $\theta$. From \eqref{Ptheta} and the expressions
in \eqref{bumptheta} one finds
\beq
\tau(p_\theta) = f_0 = \int_0^1{\rm d} x~f(x)= \theta ~.
\eeq
Furthermore, the monopole charge (i.e.\ first Chern number) of
$p_\theta$ is $1$. This is computed as the index of a Fredholm operator \cite{Co80} given by 
\begin{align}\label{topcha}
c_1(p_\theta) &:= - \frac{1}{2 \pi \ii}\,\tau \big( p_\theta \,( \partial_1 p_\theta \,\partial_2 p_\theta 
- \partial_2 p_\theta \,\partial_1p_\theta) \big) \nn \\
&\:=-6\,\int_0^1{\rm d} x~g(x)^2\,f'(x)= 1\, ,
\end{align}
with the last equality following from expression \eqref{bumptheta} of the function $f$.

When $\theta$ is irrational the projection $p_\theta$, together
with the trivial projection $1$,  generates
the $K_0$ group. The trace on $A_\theta$ gives a map
\begin{align}
\Z\,:\,K_0(A_\theta) & \quad \longrightarrow \quad \Z+\Z\,\theta \ , \nn \\
r\,[1] + m\,[p_\theta] & \quad \longmapsto  \quad \tau(1) + m\, \tau(p_\theta)  
= r + m\,\theta
\label{K0Atheta}
\end{align}
which is an isomorphism of ordered groups \cite{pva}. 
The class $m [p_\theta] $ can be represented by a Powers-Rieffel projection in the algebra itself $A_\theta$ with suitable functions in \eqref{Ptheta} of the kind \eqref{bumptheta}. 
The positive cone is the collection of (equivalence classes of) projections
with non-negative trace,
\beq
K_0^+(A_\theta)=\big\{(r,m)\in\Z^2~\big|~r+m\,\theta\geq0
\big\} \ .
\label{K0+Atheta}
\eeq

\subsection{The splitting of the algebra as a Lie bi-algebra}

We use the trace $\tau$ to define a real valued inner product on the algebra $A_\theta$:
\beq\label{innernct}
\hs{a}{b} = \textup{Im} \, \tau (ab) = \frac{1}{2 \ii} \, \sum_{m\in\Z^2} (a_m b_{-m} - a^*_{-m} b^*_{m} ). 
\eeq
This inner product is non degenerate but, as in the commutative case, it is not positive definite. 

%The sum is finite due to the condition \eqref{nctsn} on the coefficients. 

We shall denote $\u{S}_\theta$ the non-commutative torus algebra $A_\theta$ when thought of as a Lie algebra with commutator 
\eqref{cnct}. As before, we aim at using the inner product to break for a splitting $\u{S}_\theta=\u{A}\oplus \u{B}$ into 
real sub-algebras which are isotropic and paired via the inner product. We seek a splitting 
of a `purely imaginary part' (anti-hermitian operators are closed for the commutator while hermitian ones) 
and a `upper triangular' part.

It is known \cite[Thm.~3.11]{Ph06} that the ordered group ($K_0(A_\theta), K_0^+(A_\theta)$) characterizes non-commutative tori up to Rieffel-Morita equivalence: two non-commutative tori are Rieffel-Morita equivalent if and only if their ordered $K_0$-groups are isomorphic. Algebras which are equivalent in this sense are usually thought of as having the same geometry. It is then only proper  
to use the order of the $K_0$-group to label natural bases of the sub-algebras $\u{A}$ and $\u{B}$ in the splitting $\u{S}_\theta=\u{A}\oplus \u{B}$. As mentioned, 
in the limit $\theta\to 0$ the $K$-theoretical ordering tends to the lexicographic ordering we used earlier for the Lie bi-algebra of the commutative torus.

\begin{figure}
\includegraphics[width=\textwidth]{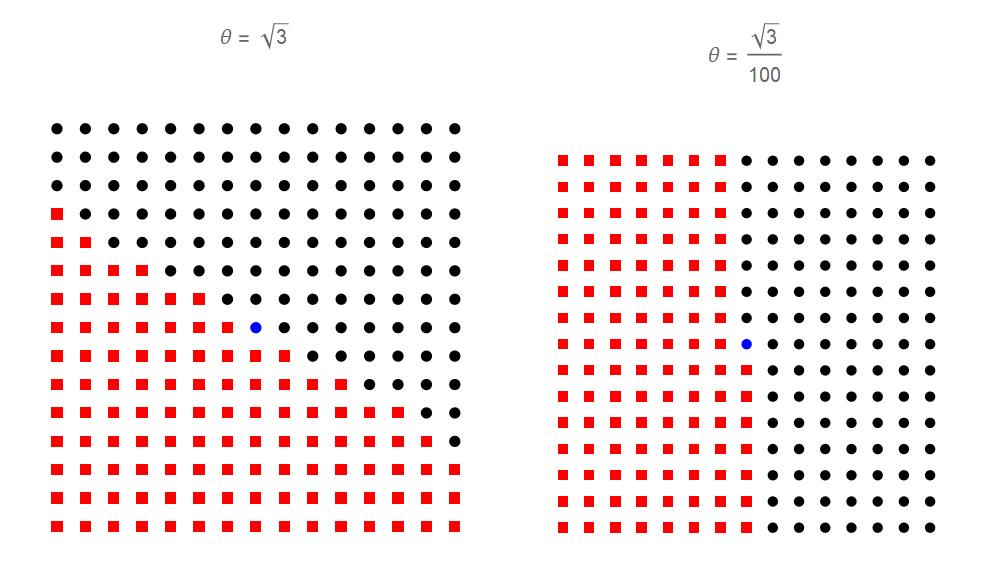}
\caption{The squares represent  elements of $(m_1,m_2)\in \mathbb{Z}^2$ with $m_1+\theta m_2<0$, while the disks represent the positive elements. In the limit $\theta\to 0$ this tends to the lexicographic ordering on $\mathbb{Z}^2$.}
\label{fig:orderedgroupZ2}
\end{figure}

Start with the real Lie sub-algebra of anti-hermitian:
$$
\u{A} = \Big\{ a = \sum\nolimits_{m \in \Z^2} a_{m} \, \hat{e}_m \, | \, a^*_{m} = - a_{-m} \Big\} ,
$$
which is clearly isotropic for the inner product \eqref{innernct}.
To avoid over counting, we use a real basis of $\u{A}$ labelled by the positive cone \eqref{K0+Atheta} of $K_0(A_\theta)$.
That is, if $m=(m_1, m_2) \in \Z$ we take
\begin{align}\label{basanct}
U_{(0,0)} & = \ii \nn \\
U_m & = \tfrac{\ii}{2} (\hat{e}_m + \hat{e}^\dagger_m) \, \qquad m_1 + m_2 \theta > 0 \nn \\ 
\wt{U}_m & = - \tfrac{1}{2} (\hat{e}_m - \hat{e}^\dagger_m)  \qquad m_1 + m_2 \theta > 0 \, .
\end{align} 
In the limit $\theta\to 0$ this tends to the lexicographic ordering we used earlier, as illustrated in fig.\ref{fig:orderedgroupZ2}. 

The dual real Lie sub-algebra $\u{B}$ is the real span of the basis elements
\begin{align} \label{basbnct}
T^{(0,0)} & = 1 \nn \\
T^n & = 2 \, \hat{e}_n \, \qquad n_1 + n_2 \theta > 0 \nn \\
\wt{T}^n & = 2 \ii \, \hat{e}_n  \, \qquad n_1 + n_2 \theta > 0 .
\end{align} 

For the inner product \eqref{innernct}
the basis \eqref{basanct} is isotropic and is dual to the basis \eqref{basbnct} which is isotropic as well, as it can be checked directly.  
The only non vanishing pairings are:
\beq\label{parfunc}
\hs{T^m}{U_m} = 1 = \hs{\wt{T}^m}{\wt{U}_m} \, .
\eeq

Next, we compute the structure constants of both the Lie algebras $A$ and $B$ starting from the commutators \eqref{cnct}.  
For $m, n \in \Z$, let us use the short notation 
\beq
s(m,n) = \sin ( \pi \theta \, m \wedge n)
\eeq
(an odd function on each argument) and the convention $m > n$ if 
and only if $m_1 + m_2 \theta > n_1 + n_2 \theta$.

For the algebra $\u{A}$: $[X_{a},X_{b}]=\Gamma_{ab}^{c}X_{c}$, 
the only non vanishing $\Gamma$'s are computed to be
\begin{align}
\Gamma^{m+n}_{m,n}  &=  - s(m,n), \qquad \Gamma^{\wt{m+n}}_{m,\wt{n}} =  - s(m,n), 
\qquad \Gamma^{m+n}_{\wt{m},\wt{n}} = s(m,n), \nn \\
\Gamma^{m-n}_{m,n}  &= s(m,n), \qquad \Gamma^{\wt{m-n}}_{m,\wt{n}} =  - s(m,n), 
\qquad \Gamma^{m-n}_{\wt{m},\wt{n}} = s(m,n), \qquad \textup{for} \quad m > n , \nn \\
\Gamma^{n-m}_{m,n}  &= s(m,n), \qquad \Gamma^{\wt{n-m}}_{m,\wt{n}} =  - s(m,n), 
\qquad \Gamma^{n-m}_{\wt{m},\wt{n}} = s(m,n), \qquad \textup{for} \quad m < n ,
\end{align}
and their antisymmetric ones in the two lower indices.

For the algebra $\u{B}$: $[X^{a},X^{b}]=\Delta_{c}^{ab}X_{c}$, the only non vanishing $\Delta$'s are computed to be
\beq
\Delta^{m,n}_{\wt{m+ n}} = 4 s(m,n) , \qquad \Delta^{m,\wt{n}}_{m+ n} = - 4 s(m,n), 
\qquad \Delta^{\wt{m},\wt{n}}_{\wt{m+ n}} = -4 s(m,n)
\eeq
and their antisymmetric ones in the two upper indices.

Finally, for the mixed compatible ones $[X^{a},X_{b}]=\Gamma_{bd}^{a}X^{d}-\Delta_{b}^{ad}X_{d}$, one computes

\begin{align}
[T^m, U_n] = 
\begin{cases}
\, \Gamma^{m}_{n,m+n} \, T^{m+n} + \Gamma^{m}_{n,m-n} \, T^{m-n}  \qquad & m > n \\
\, \Gamma^{m}_{n,m+n} \, T^{m+n} + \Gamma^{m}_{n,n-m} \, T^{n-m} - \Delta^{m,\wt{n-m}}_{n} \wt{U}_{n-m} \qquad & m <  n 
\end{cases}
\end{align}

\begin{align}
[T^m, \wt{U}_n] = 
\begin{cases}
\, \Gamma^{m}_{\wt{n},\wt{m+n}} \, \wt{T}^{m+n} + \Gamma^{m}_{\wt{n},\wt{m-n}} \, \wt{T}^{m-n}  \qquad & m > n \\
\, \Gamma^{m}_{\wt{n},\wt{m+n}} \, \wt{T}^{m+n} + \Gamma^{m}_{\wt{n},\wt{n-m}} \, 
\wt{T}^{n-m} - \Delta^{m, n-m}_{\wt{n}} U_{n-m} \qquad & m < n 
\end{cases}
\end{align}

\begin{align}
[\wt{T}^m, U_n] = 
\begin{cases}
\, \Gamma^{\wt{m}}_{n,\wt{m+n}} \, \wt{T}^{m+n} + \Gamma^{\wt{m}}_{n,\wt{m-n}} \, \wt{T}^{m-n}  \qquad & m > n \\
\, \Gamma^{\wt{m}}_{n,\wt{m+n}} \, \wt{T}^{m+n} + \Gamma^{\wt{m}}_{n,\wt{n-m}} \, 
\wt{T}^{n-m} - \Delta^{\wt{m},n-m}_{n} U_{n-m} \qquad & m <  n 
\end{cases}
\end{align}

\begin{align}
[\wt{T}^m, U_n] = 
\begin{cases}
\, \Gamma^{\wt{m}}_{\wt{n}, m+n} \, T^{m+n} + \Gamma^{\wt{m}}_{\wt{n}, m-n} \, T^{m-n}  \qquad & m > n \\
\, \Gamma^{\wt{m}}_{\wt{n}, m+n} \, T^{m+n} + \Gamma^{\wt{m}}_{\wt{n}, n-m} \, 
T^{n-m} - \Delta^{\wt{m}, \wt{n-m}}_{\wt{n}} \wt{U}_{n-m} \qquad & m <  n 
\end{cases}
\end{align}

The above shows the Lie-bi-algebra structure of the non-commutative torus $\u{S}_\theta=\u{A}\oplus \u{B}$.

\appendix

\section{The pairings: rational case}\label{app1}

We need the scalar product between $E$'s in \eqref{gencs} and $F$'s in \eqref{eff},  \eqref{efft}. One finds explicitly
\begin{align}
\tr (f_{a,b} \, e^*_{r,s}) & = \delta(a-r)\, \omega^{- \frac{1}{2}a(b-s)} \sum_{n=a}^{N-1} \omega^{n(b-s)} \nn \\
 \tr (\wt{f}_{a,b} \, e_{r,s}) & = \delta(a-r)\, \omega^{ \frac{1}{2}a(b-s)} \sum_{n=0}^{a-1} \omega^{-n(b-s)} \label{r=a}
 \end{align}

\begin{align}
\tr (f_{a,b} \, e_{r,s}) & = \delta(a+r)\, \omega^{ - \frac{1}{2}a(b+s)} \sum_{n=a}^{N-1} \omega^{n(b+s)} \nn \\
 \tr (\wt{f}_{a,b} \, e^*_{r,s}) & = \delta(a+r)\, \omega^{ \frac{1}{2}a(b+s)}  \sum_{n=0}^{a-1}  \omega^{-n(b+s)} \label{r=-a}
 \end{align}
Putting these together, and using $\sum_{j=0}^{N-1} \omega^{jm}= N \delta(m)$, we get:

\medskip
\noindent
For $a=r$:
\begin{align}
\tr (f_{a,b} \, e^*_{r,s} + \wt{f}_{a,b} \, e_{r,s} ) & = \omega^{- \frac{1}{2}a(b-s)} \sum_{n=a}^{N-1} \omega^{n(b-s)} + \omega^{ \frac{1}{2}a(b-s)} \sum_{n=0}^{a-1} \omega^{-n(b-s)}  \label{ar+} \\
& =
\begin{cases} N \quad & b=s  \\
- \omega^{- \frac{1}{2}a(b-s)} \sum_{n=0}^{a-1} \omega^{n(b-s)} + \omega^{ \frac{1}{2}a(b-s)} \sum_{n=0}^{a-1} \omega^{-n(b-s)} \in \ii \R \quad & b\not=s \nn
\end{cases}
 \end{align}

\begin{align}
\tr (f_{a,b} \, e^*_{r,s} - \wt{f}_{a,b} \, e_{r,s} ) & = \omega^{- \frac{1}{2}a(b-s)} \sum_{n=a}^{N-1} \omega^{n(b-s)} - \omega^{ \frac{1}{2}a(b-s)} \sum_{n=0}^{a-1} \omega^{-n(b-s)} \label{ar-} \\
& =
\begin{cases} N -2a \quad & b=s \\
- \omega^{- \frac{1}{2}a(b-s)} \sum_{n=0}^{a-1} \omega^{n(b-s)} - \omega^{ \frac{1}{2}a(b-s)} \sum_{n=0}^{a-1} \omega^{-n(b-s)} \in \R \quad  & b\not=s \nn
\end{cases}
 \end{align}

\medskip
\noindent
For $a-r$:
\begin{align}
\tr (f_{a,b} \, e_{r,s} + \wt{f}_{a,b} \, e^*_{r,s} ) & = \omega^{- \frac{1}{2}a(b+s)} \sum_{n=a}^{N-1} \omega^{n(b+s)} + \omega^{ \frac{1}{2}a(b+s)} \sum_{n=0}^{a-1} \omega^{-n(b+s)} \label{a-r+}\\
& =
\begin{cases}  N \quad & b=-s \\
- \omega^{- \frac{1}{2}a(b+s)} \sum_{n=0}^{a-1} \omega^{n(b+s)} + \omega^{ \frac{1}{2}a(b+s)} 
\sum_{n=0}^{a-1} \omega^{-n(b+s)} \in \ii \R \quad & b\not= -s \nn
\end{cases}
 \end{align}

\begin{align}
\tr (f_{a,b} \, e_{r,s} - \wt{f}_{a,b} \, e^*_{r,s} ) & = \omega^{- \frac{1}{2}a(b+s)} \sum_{n=a}^{N-1} \omega^{n(b+s)} - \omega^{ \frac{1}{2}a(b+s)} \sum_{n=0}^{a-1} \omega^{-n(b+s)} \label{a-r-} \\
& =
\begin{cases} N - 2a \quad & b=-s \\
- \omega^{- \frac{1}{2}a(b+s)} \sum_{n=0}^{a-1} \omega^{n(b+s)} - \omega^{ \frac{1}{2}a(b+s)} \sum_{n=0}^{a-1} \omega^{-n(b+s)} \in \R \quad  & b\not= -s \nn
\end{cases}
 \end{align}
 
 \noindent
From \eqref{r=a} and \eqref{r=-a} the pairing of a $T^{a,b}$  
(either without or with a tilde) and a $T^{a,b}$ (again either without or with a tilde) 
is zero unless $r=a$ or $r=-a$. Then, on the one end, from \eqref{ar-} and \eqref{a-r-},
for any $s$, 
$$
\tr (\wt{T}^{a,b} \, U_{\pm a, s}) \in \R \qquad \tr (T^{a,b} \, \wt{U}_{\pm a, s}) \in \R
$$
and these lead for any $s$ to
$$
\hs{\wt{T}^{a,b}}{U_{\pm a, s}} = \hs{T^{a,b}}{\wt{U}_{\pm a, s}} = 0 
$$
On the other end, from \eqref{ar+} we get
$$
\tr (T^{a,b} \, U_{a, b}) = \ii  \qquad \tr (T^{a,b} \, U_{a, s}) \in \R  \quad s\not=b
$$
$$
\tr (\wt{T}^{a,b} \, \wt{U}_{a, b}) = \ii  \qquad \tr (\wt{T}^{a,b} \, \wt{U}_{a, s}) \in \R  \quad s\not=b
$$
that is the only non vanishing pairings are 
\beq\label{parnot0}
\hs{T^{a,b}}{U_{a, b}} = 1 = \hs{\wt{T}^{a,b}}{\wt{U}_{a, b}} 
\eeq
In particular for the diagonal matrices, $\ft$ and $\ii \ft$ are isotropic and dually paired with pairings:
\beq\label{diag0}
\hs{H_A}{\ii H_B} = \delta_{A,B} \qquad  A = a, \tilde{a}, \, \, B = b, \tilde{b}
\eeq
with the ranges of the indices as in \eqref{bcar}.

\section{Taft algebras and their action}\label{sec:Taft}
Let $\omega(=e^{2 \pi \ii / N})$ be a primitive $N$-the root of unity. 
The \emph{Taft algebra} $T_{N}$, introduced in \cite{Taft}, 
is a Hopf  algebra which is neither commutative nor co-commutative. Firstly, $T_{N}$
is the $N^{2}$-dimensional unital algebra generated 
by generators $R$, $G$ subject to the relations:
$$
R^{N}=0\, , \quad G^{N}=1\, , \quad R G - \omega \, G R=0 \, .
$$
It is a Hopf algebra with coproduct: 
\begin{align*}
\Delta(R):=1\otimes R+R\otimes G, \qquad \Delta(G):=G \otimes G \, ;
\end{align*}
 counit: $\varepsilon(R):=0$, $\varepsilon(G):=1$, 
and antipode: $S(R): = - R G^{-1}$, $S(Q) := G^{-1}$.
The four dimensional algebra $T_2$ is also known as the \emph{Sweedler algebra}.

For any $s\in \C$, let $A_{s}$ be the unital algebra generated by elements $r, g$ with relations: 
$$
r^{N}=s\, , \quad g^{N} = 1\, , \quad r g - \omega \, g r = 0\, . 
$$
When $s=1$ this is just the algebra of rational non-commutative torus of Section \ref{rnct}.

\noindent
The algebra $A_{s}$ is a right $T_{N}$-comodule algebra, with coaction $\delta^A : A_s \to A_s \otimes T_N$ defined by
\begin{align}\label{coTaft}
\delta^A(r):=1\otimes R+ r \otimes G,  \qquad \delta^A(g):=g \otimes G.
\end{align}
The algebra of corresponding coinvariant (invariant for the coaction) elements, that is elements $x\in A_s$ such that $\delta^A(x) = x \otimes 1$ is just the algebra $\C$. Moreover, the canonical map,
$$
\chi : A_s \otimes A_s \to A_s \otimes T_N, \quad \chi(x \otimes y) = (x \otimes 1) \, \delta^A(y) 
$$ 
is an isomorphism. 
This states that the ``coaction is free and transitive" and the extension $\C = (A_{s})^{T_n} \subset A_s$ is a non-commutative principal bundle over a point $\{ * \}$ whose algebra of function is the coinvariant algebra $\C$. 
One also says that $A_{s}$ is a $T_{N}$-Galois object. 

Contrary to the commutative case, these are not trivial.
It is known (see  \cite{Ma94}, Prop. 2.17 and Prop. 2.22) that any $T_{N}$-Galois object is isomorphic to $A_s$ for some $s\in \C$ and that any two such Galois objects $A_s$ and $A_t$ are isomorphic if and only if $s=t$. Thus the equivalence classes of  $T_{N}$-Galois objects are in bijective correspondence with the abelian group $\C$. The translation map of the coaction, $\tau := (\chi^{-1})_{| 1 \otimes T_N} : T_N \to A_s \otimes A_s$, is given on generators by
\beq\label{trTaft}
\tau(G) = g^{-1} \otimes g , \quad 
\tau(R) = 1 \otimes r - r g^{-1} \otimes g .
\eeq

\end{document}